\documentclass[a4paper, 10pt, conference]{IEEEtran}
\IEEEoverridecommandlockouts
\usepackage{cite}
\usepackage{amsmath,amssymb,amsfonts}
\usepackage{algorithmic}
\usepackage{graphicx}
\usepackage{textcomp}
\usepackage{xcolor}
\usepackage{comment}
\usepackage{csquotes}
\usepackage{hyperref}
\usepackage{tikz}
\usetikzlibrary{positioning}

\makeatletter 
\newcommand{\linebreakand}{%
  \end{@IEEEauthorhalign}
  \hfill\mbox{}\par
  \mbox{}\hfill\begin{@IEEEauthorhalign}
}
\makeatother 

\def\BibTeX{{\rm B\kern-.05em{\sc i\kern-.025em b}\kern-.08em
    T\kern-.1667em\lower.7ex\hbox{E}\kern-.125emX}}
\begin{document}

\title{Comparing Continuous and Retrospective Emotion Ratings in Remote VR Study}


\author{
\IEEEauthorblockN{Maximilian Warsinke}
\IEEEauthorblockA{\textit{Quality and Usability Lab} \\
\textit{Technische Universität Berlin}\\
Berlin, Germany \\
warsinke@tu-berlin.de}
\and
\IEEEauthorblockN{Tanja Kojić }
\IEEEauthorblockA{\textit{Quality and Usability Lab} \\
\textit{Technische Universität Berlin}\\
Berlin, Germany \\
tanja.kojic@tu-berlin.de}
\and
\IEEEauthorblockN{Maurizio Vergari}
\IEEEauthorblockA{\textit{Quality and Usability Lab} \\
\textit{Technische Universität Berlin}\\
Berlin, Germany \\
maurizio.vergari@tu-berlin.de}
\linebreakand
\IEEEauthorblockN{Robert Spang}
\IEEEauthorblockA{\textit{Quality and Usability Lab} \\
\textit{Technische Universität Berlin}\\
Berlin, Germany \\
spang@tu-berlin.de}
\and
\IEEEauthorblockN{Jan-Niklas Voigt-Antons}
\IEEEauthorblockA{\textit{Immersive Reality Lab} \\
\textit{Hochschule Hamm-Lippstadt}\\
Lippstadt, Germany \\
jan-niklas.voigt-antons@hshl.de}
\and
\IEEEauthorblockN{Sebastian Möller}
\IEEEauthorblockA{\textit{Quality and Usability Lab} \\
\textit{Technische Universität Berlin and DFKI}\\
Berlin, Germany \\
sebastian.moeller@tu-berlin.de}
}

\maketitle

\begin{abstract}
This study investigates the feasibility of remote virtual reality (VR) studies conducted at home using VR headsets and video conferencing by deploying an experiment on emotion ratings. 20 participants used head-mounted displays to immerse themselves in 360° videos selected to evoke emotional responses. The research compares continuous ratings using a graphical interface to retrospective questionnaires on a digitized Likert Scale for measuring arousal and valence, both based on the self-assessment manikin (SAM). It was hypothesized that the two different rating methods would lead to significantly different values for both valence and arousal. The goal was to investigate whether continuous ratings during the experience would better reflect users' emotions compared to the post-questionnaire by mitigating biases such as the peak-end rule. The results show significant differences with moderate to strong effect sizes for valence and no significant differences for arousal with low to moderate effect sizes. This indicates the need for further investigation of the methods used to assess emotion ratings in VR studies. Overall, this study is an example of a remotely conducted VR experiment, offering insights into methods for emotion elicitation in VR by varying the timing and interface of the rating.
\end{abstract}

\newcommand\copyrighttext{%
  \footnotesize \textcopyright\ 2024 IEEE. Personal use of this material is permitted. Permission from IEEE must be obtained for all other uses, in any current or future media, including reprinting/republishing this material for advertising or promotional purposes, creating new collective works, for resale or redistribution to servers or lists, or reuse of any copyrighted component of this work in other works. The original version of this article is available at: \href{https://ieeexplore.ieee.org/document/10598301}{https://ieeexplore.ieee.org/document/10598301} DOI:\href{https://doi.org/10.1109/QoMEX61742.2024.10598301}{10.1109/QoMEX61742.2024.10598301}%
}

\newcommand\copyrightnotice{%
\begin{tikzpicture}[remember picture,overlay,shift={(current page.south)}]
  \node[anchor=south,yshift=10pt] at (0,0) {\fbox{\parbox{\dimexpr\textwidth-\fboxsep-\fboxrule\relax}{\copyrighttext}}};
\end{tikzpicture}%
}

\copyrightnotice


\section{Introduction}

VR technology has received substantial attention from academics over the last several years \cite{360versus3d}. Many of these research activities have typically been carried out inside a controlled laboratory environment, with researchers and participants located near one another \cite{kourtesis2020guidelines}. The appearance of the COVID-19 epidemic has highlighted the importance of investigating and putting into practice various distant recruiting tactics.

While conducting VR studies in the laboratory has provided useful insights into user behavior, preferences, and usability of VR apps, it is not without limits. One factor to consider is the possibility of a bias introduced by the testing environment. In the laboratory, the physical presence of researchers, moderators, and observers may impact participants' emotions, thereby skewing the results \cite{marsh2018user}. This \enquote{observer effect} could be even more noticeable in VR, where the presence of other people can affect sensation, immersion, and overall experience.

Therefore, remote testing allows for VR trials with participants in their homes or chosen locations, while connecting to researchers through web video streaming. Furthermore, remote testing allows one to interact with participants in environments in which they are most comfortable, such as their native context \cite{saffo2021remote}. This scenario is useful when researching real-world VR applications, such as remote work collaboration tools, virtual schools, and therapeutic interventions, and investigating various research problems, from usability and user experience to cognitive and emotional reactions in VR.

Some efforts have already been reported, such as the research and investigation of the potential benefits of VR and extended reality (XR) technology in remote settings. These include investigating the use of data collection capabilities built into head-mounted displays (HMDs), such as hand and gaze tracking, and the benefits of mobility and repeatability provided by XR in experimental settings \cite{ratcliffe2021extended}. Despite their usefulness in other domains, the VR research community is careful to embrace remote engagement methods \cite{preece2016citizen}. 

To begin such research based on remote testing, an experiment that would normally take place in a laboratory was conducted. The experiment investigates the evaluation of emotions after and during viewing of 360° videos in VR as an approach to get closer to the user's \enquote{real} emotions. The values of valence and arousal according to the circumplex model of effect \cite{posner2005circumplex} were chosen as the self-report measurements. In this study, two rating methods were developed that differ in interface and timing, both based on the SAM rating scale \cite{bradley_measuring_1994}. The continuous rating method requires participants to rate their emotions multiple times on a 2D interface while 360° videos are running in the background. The retrospective rating method is a digitized Likert scale that is incorporated as a floating interface into the VR environment and is shown after the video ends. The following hypotheses were formulated for this experiment:

\vspace{14pt}

\textbf{Hypothesis 1 (\(H_1\)):}
There is a statistically significant difference in valence ratings between participants' ratings from the continuous and retrospective rating methods when watching 360° videos in VR.

\textbf{Hypothesis 2 (\(H_2\)):}
There is a statistically significant difference in arousal ratings between participants' ratings from the continuous and retrospective rating methods when watching 360° videos in VR.

\vspace{14pt}

\section{Related Work}

\subsection{Remote VR-studies}
VR devices' popularity in private households has increased in the last couple of years with ownership rates reaching 12\% in the United States by the year 2022 \cite{deloitte}. Researchers suggest conducting remote VR studies, reporting reliable results compared to results from the lab \cite{vr_outside_lab}. Using participants-owned HMDs could increase the sample sizes of experiments with the disadvantage of drawing from a more homogeneous group, consisting of proficient VR users \cite{saffo2021remote}. Alternatively, delivering VR setups to the homes can foster diverse participants with the risk of insufficient technical knowledge to set up the VR devices \cite{Canyousetitup}. Mottelson et al. reported a diverse and high number of participants at lower costs in two case studies on unsupervised remote experiments \cite{unsupervised_vr}. A supervised remote study with a scheduled Zoom call has been explored by Mathis et al. who describe the potential to complement conventional experiments done in the laboratory \cite{stay_home}.


\subsection{Inducing Emotion in VR}
Using virtual reality experiences to induce human emotions is a widely used approach in emotion research and human-computer interaction (HCI) \cite{emotion-elicitation-review, survey_on_cognitive_affective}. Common practices utilize 360° videos, 3D environments\cite{realtime_vr_annotations}, and VR games \cite{dontworrybehappy} as stimuli to evoke emotional reactions in experiments.

While three-dimensional VR environments appear to induce a higher feeling of presence \cite{360versus3d}, 360° videos seem to be an efficient alternative to use as stimuli for experiments, because of higher online availability and fewer development efforts compared to 3D content \cite{brivio_virtual_2021}. Additionally, while current advancements in 3D modeling allow for a realistic depiction of various environments, creating human characters or animals is more challenging, causing unintended responses such as an uncanny valley effect \cite{mori2012uncanny}. By contrast, 360° videos can offer a more authentic experience by capturing real humans and animals in real-world scenes using 360° cameras. For this purpose, a database of 360° videos has been collected and annotated \cite{li_public_2017}. The database features full spherical view footage of diverse situations that can be explored naturally using an HMD.

Established methods for assessing emotions include quantitative and qualitative approaches that use physiological measurements and questionnaires. Physiological measurements include electroencephalography (EEG)\cite{Pinilla_Voigt-Antons_Garcia_Raffe_Moller_2023}, skin conductance capturing \cite{PEM360}, and eye tracking \cite{vreed}. Novel studies have used head movement during experiences to successfully predict emotional states \cite{headmovement, headmovement2}. Subjective emotion ratings are commonly reported as valence and arousal following the circumplex model of effect \cite{Russell_1980}. To gather data, post-experience questionnaires such as SAM have been employed \cite{vr_sam_questionnaire}. This self-assessment questionnaire uses symbolic depictions of emotions combined with a Likert scale for appraisal. This led to the further development of graphical rating interfaces like the \enquote{EmojiGrid} \cite{emojigrid_2D} developed by Toet et al. which was later used for emotion elicitation of 360° videos in VR \cite{emojigrid_vr_basevalues}. Continuous ratings during the experience have been explored to observe a post-experience effect on emotion estimation without showing significant differences \cite{voigt-antons_comparing_2020}. When working with self-reports, psychological biases must be taken into account, such as the peak-end rule, which describes an overestimation of an effect when a situation is evaluated retrospectively \cite{Fredrickson_Kahneman_1993}.

\section{Methods}
\subsection{Experimental Design}
The experimental design was used to fit a previous study\cite{voigt-antons_comparing_2020}. Therefore, a within-subject design was chosen such that every participant gave ratings for both rating methods, eliminating the influence of individual variability. The sequence of the rating method was alternated for every participant to counterbalance the possible sequence effects. As a questionnaire,  the self-assessment manikin (SAM) was chosen to assess emotions to ensure comparability with previous research. During the experiment, the answers to the rating scales were saved on the device's storage and could later be extracted for analysis.

\subsection{Participants}
A total of 21 participants between the ages of 20 and 29 took part in the experiment, including 18 men, 2 women and 1 non-binary person. Due to incomplete ratings, data from one participant were excluded from the dataset, resulting in 20 participants for the analysis. All the remaining participants were students, apart from one who worked as a Python developer. Their age ranged from 20 to 29. Among the 7 participants who self-reported having vision impairments, 6 wore glasses and one wore lenses. Regarding the experience with the VR system, the distribution was as follows: 9 participants used a VR system 1 to 2 times, 6 participants used it 3 to 5 times, 1 participant used it 6 to 9 times, and 4 participants used it 10 or more times. The mean Affinity for Technology Interaction Score (ATI) \cite{franke2019personal} of the participants across all items was 4.76 ($SD = 0.50$).

\begin{table}[ht]
\centering
\caption{List of all videos with their ID from Li et al. \cite{li_public_2017}.}
\begin{tabular}{|l|l|c|c|}
\hline
\textbf{ID} & \textbf{Name} & \textbf{Slice} & \textbf{Quadrant}\\
\hline
68 & Jailbreak 360 & 2:39 – 3:39 & LVHA\\
20 & War Knows no Nation 1 & 4:44 – 5:44 & LVHA \\
20 & War Knows no Nation 2 & 3:14 – 4:14 & LVHA\\
21 & Zombie Apocalypse Horror & 0:40 – 1:40 & LVHA\\
\hline
63 & Nasa Rocket Launch & 3:15 – 4:15 & HVHA\\
50 & Puppies 360 & 0:04 – 1:04 & HVHA \\
64 & Surrounded by Elephants & 0:30 – 1:30 & HVHA\\
69 & Walk the Tight Rope & 0:27 – 1:27 & HVHA\\
\hline
8 & Happyland 360 & 2:00 – 3:00 & LVLA \\
18 & The Displaced 1 & 2:23 – 3:23 & LVLA \\
18 & The Displaced 2 & 3:35 – 4:35 & LVLA\\
16 & Solitary Confinement & 0:00 – 1:00 & LVLA \\
\hline
43 & Alice the Baby & 0:00 – 1:00 & HVLA\\
26 & Getting licked by a Cow & 0:00 – 1:00 & HVLA \\
22 & Great Ocean Road & 0:00 – 1:00 & HVLA\\
32 & Malaekahana Sunrise & 1:20 – 2:20 & HVLA\\
\hline
\end{tabular}
\label{table:table_all_values}
\end{table}

\subsection{Stimuli}
The 360° clips used in the experiment were selected from the Stanford public database\footnote[1]{https://vhil.stanford.edu/public-database-360-videos-corresponding-ratings-arousal-and-valence} that consist of 73 videos in total with their corresponding valence and arousal values. Video materials are mostly publicly available on popular websites, such as YouTube \cite{li_public_2017}. The clips used were selected based on their baseline values. To ensure equal distribution of emotions, 4 clips were used from each quadrant of the circumplex model of affect \cite{posner2005circumplex}.
\vspace{11pt}
\begin{itemize}
    \item \textbf{LVHA} (Low valence, high arousal)
    \item \textbf{HVHA} (High valence, high arousal)
    \item \textbf{LVLA} (Low valence, low arousal)
    \item \textbf{HVLA} (High valence, low arousal)
\end{itemize}
\vspace{12pt}
Each video was shortened to a duration of 60 seconds to prevent differences in results arising from varying lengths. This consistency is particularly crucial for retrospective ratings, in which longer video clips may lead to variations in the review of emotions. Due to a lack of video clips from the quadrant with low valence and high arousal (LVHA), two clips were taken from the same source video, depicting different scenes from the video. The same procedure was repeated for low-valence and low-arousal quadrants (LVLA). The complete list of video clips used is provided in Table \ref{table:table_all_values} with their respective quadrants.

\subsection{Procedure}
The experiment was conducted and supervised in a remote setting and approved by the local ethics committee of the Technical University of Berlin. Participants received the VR-Setup with the pre-installed experiment application via parcel delivery and then showed up to a scheduled appointment to meet on Zoom with the instructor. The study was conducted using a Meta Quest HMD. The procedure was explained, and the participants digitally signed a consent form. Afterward, the participants were given an online questionnaire on demographics and 9 items from the Affinity for Technology Interaction (ATI) questionnaire. The participants were then requested to cast from the Meta Quest to their computer and share their screen so that the instructor could retrace the application. Participants were then allowed to start the experiment in a practice mode in which they could get familiar with the HMD and its controls. Additionally, an explanation of the SAM and the concepts of valence and arousal were given. During the main part of the experiment, the participants watched and rated 4 randomly selected 360° videos (1 from each quadrant). The ratings were either given retrospectively or continuously as predetermined for each participant. Subsequently, another 4 randomly selected videos (1 from each quadrant) were chosen for the alternative rating method. Once participants watched 8 videos, they were asked to complete the final questionnaire online, which asked for a subjective evaluation of the rating methods and gave participants the opportunity to leave feedback. After debriefing, the Zoom call was closed.

\subsection{Rating Methods}
\subsubsection{Retrospective}
\begin{figure}[ht]
\centering
\includegraphics[width=0.45\textwidth]{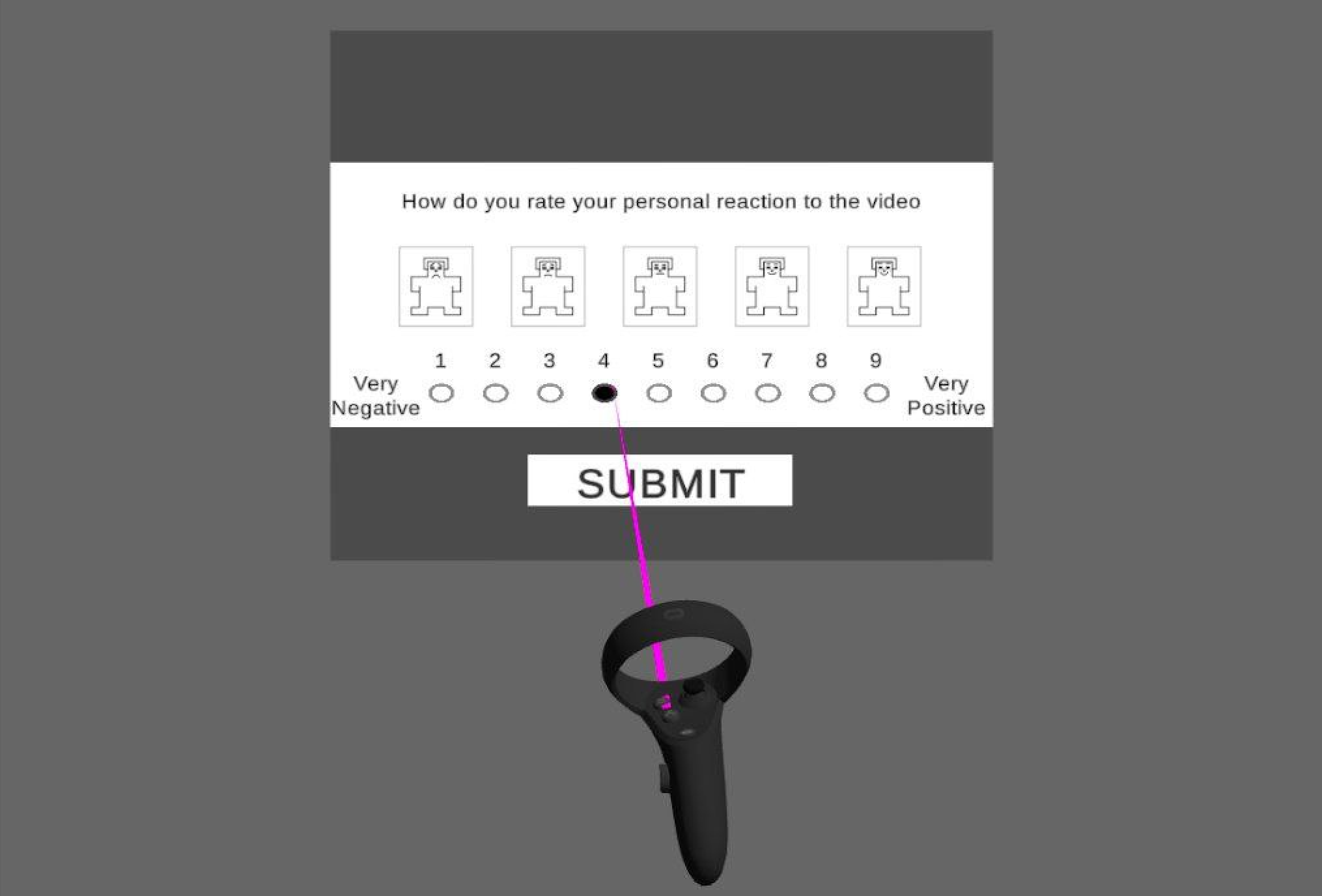}
\caption{Retrospective rating interface for valence with a digitized Likert scale based on the SAM.}
\label{fig:fig_retrospective}
\vspace{-1em}
\end{figure}

In the retrospective rating method, the participants first watched the video entirely. They were then confronted with a post-stimulus interface, where they could rate by pointing and selecting with the controller and confirming it with a submit button (Fig.  \ref{fig:fig_retrospective}). Valence and arousal ratings were collected successively using a digitized version of the SAM, including the 9-point Likert scale and the five corresponding pictures \cite{bradley_measuring_1994}. 

\subsubsection{Continuous}
\begin{figure}[ht]
\centering
\includegraphics[width=0.45\textwidth]{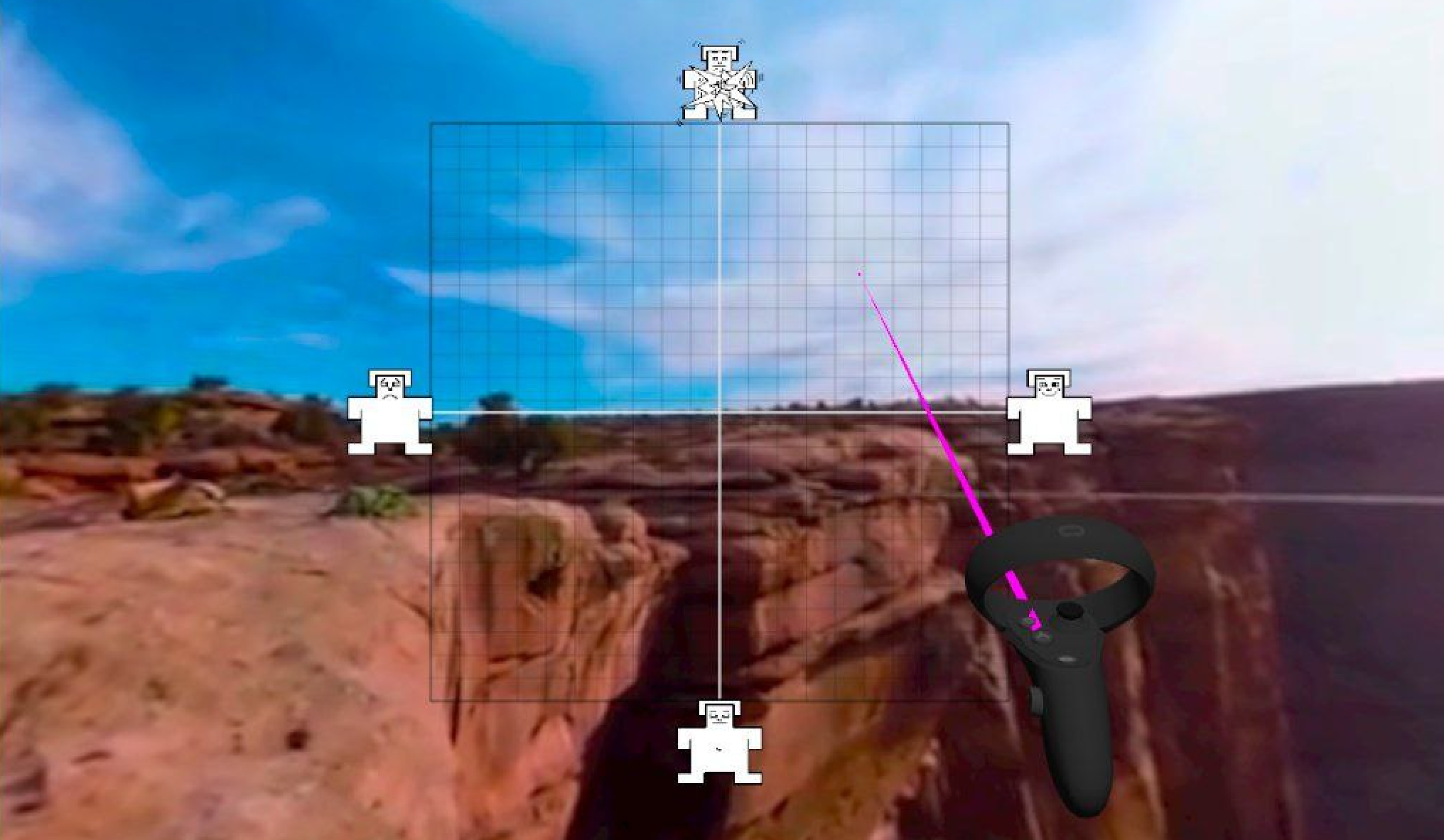}
\caption{Continuous 2D-overlay during the video based on the SAM with the x-axis denoting valence and the y-axis denoting arousal.}
\label{fig:fig_continuous}
\vspace{-1em}
\end{figure}

For the continuous rating method, a two-dimensional grid was overlaid onto the 360° videos (Fig. \ref{fig:fig_continuous}), where participants could place their ratings using the Meta Quest controller. The axes on the grid represent valence and arousal and include the pictograms of the original SAM questionnaire. Ratings were given after the play of an auditory cue that occurred every 10 seconds, resulting in 5 ratings for the 60-second duration of the video. When rating, the video continued to play in the background with no interruption, which resulted in non-mandatory voting. If a participant did not provide a rating after the cue, the data point was skipped. The ratings of each video were then averaged to get a single value for both valence and arousal.

\subsection{Statistical Testing}
The values from the continuous feedback interface were on a 2-dimensional scale with valence and arousal values from $-$50 to 50. These values were converted using simple linear scaling to a 1 to 9 scale to match the values from the retrospective SAM scale (i.e., $-$50 corresponding to 1 and 50 corresponding to 9), thus making a comparison possible.
With 20 participants watching 8 video clips each, a total of 160 ratings were collected for both arousal (A) and valence (V). These ratings are equally distributed between both rating methods, retrospective ($V_{\text{Retro}}$, $A_{\text{Retro}}$), and continuous ($V_{\text{Cont}}$, $A_{\text{Cont}}$), leaving 4 groups of values, each with 80 ratings. For statistical testing, these ratings were divided into groups according to the 4 quadrants (LVHA, HVHA, LVLA, HVLA) to maintain the influence of the baseline values of the stimuli while ensuring statistical power. This results in a sample size of 20 data points for each group, which were compared to assess the differences in rating methods. 

Eight test procedures were conducted to compare the means between the rating methods, all following a consistent protocol. The preferred statistical test was the paired $t$-test, which relies on the assumptions of homogeneity and normal distribution. Homogeneity was assessed using Levene's test, and normal distribution was evaluated using the Shapiro-Wilk test. If at least one of these assumptions was violated, a non-parametric test, specifically the Wilcoxon signed-rank test, was employed; otherwise, the paired $t$-test was considered suitable. The effect sizes were evaluated using Cohen's $d$ for pairs with a normal distribution and the Wilcoxon effect size $r$ for non-normally distributed pairs.

To compare the rating data with the baseline values, instead of grouping the values by quadrant, the rating means for each video were calculated. To increase comparability, for this analysis, only the retrospective scores were considered, as they were assessed in a similar way, namely retrospectively and using a 9-point Likert scale. Spearman’s rank correlation coefficients or Pearson’s correlation coefficients were selected depending on whether the assumptions of normal distribution were met.


\section{Results}

\subsection{Correlation to Baseline Values}

The mean retrospective arousal and mean retrospective valence values for each video were calculated to assess Pearson's correlation coefficients with the baseline values from Li et al. \cite{li_public_2017}. The analysis revealed a strongly positive correlation ($r=0.816, p<0.001$) between retrospective valence ratings and baseline valence values. A moderately to strong positive correlation ($r=0.668, p=0.003$) was observed between retrospective arousal ratings and baseline arousal values.


\begin{table*}[ht]
  \centering
  \caption{Statistical tests comparing valence for retrospective rating and continuous rating}
  \begin{tabular}{|l|c|c|c|c|c|c|c|c|}
    \hline
    Group & Mean $V_{\text{Cont}}$ & Mean $V_{\text{Retro}}$ & SD $V_{\text{Cont}}$ & SD $V_{\text{Retro}}$ & Test Used & Test Statistic & $p$-value & Effect Size\\
    \hline
    LVHA & 4.64 & 4.00 & 0.81 & 1.81 & Wilcoxon signed-rank & 1.71 & 0.086 & 0.27 \\
    HVHA & 5.01 & 6.25 & 0.44 & 1.77 & Wilcoxon signed-rank & $-$2.61 & \textbf{0.009*} & $-$0.41 \\
    LVLA & 4.71 & 3.52 & 0.94 & 1.19 & Paired $t$-test & 3.84 & \textbf{0.001*} & 1.11 \\
    HVLA & 5.63 & 6.80 & 1.09 & 1.94 & Wilcoxon signed-rank & $-$2.17 & \textbf{0.030*} & $-$0.34 \\
    \hline
  \end{tabular}
  \label{table:valence_tests}
\end{table*}

\begin{table*}[ht]
  \centering
  \caption{Statistical tests comparing arousal for retrospective rating and continuous rating}
  \begin{tabular}{|l|c|c|c|c|c|c|c|c|}
    \hline
    Group & Mean $A_{\text{Cont}}$ & Mean $A_{\text{Retro}}$ & SD $A_{\text{Cont}}$ & SD $A_{\text{Retro}}$ & Test Used & Test Statistic & $p$-value & Effect Size \\
    \hline
    LVHA & 6.19 & 5.80 & 1.50 & 1.99 & Wilcoxon signed-rank & 1.20 & 0.232 & 0.18 \\
    HVHA & 5.69 & 5.20 & 1.72 & 2.31 & Paired $t$-test & 0.87 & 0.395 & 0.24 \\
    LVLA & 4.62 & 4.15 & 1.21 & 1.81 & Paired $t$-test & 1.38 & 0.184 & 0.30 \\
    HVLA & 4.81 & 3.90 & 1.99 & 2.02 & Paired $t$-test & 1.47 & 0.158 & 0.45 \\
    \hline
  \end{tabular}
  \label{table:arousal_tests}
\end{table*}

\begin{figure*}[ht]
    \centering
    \small
    \begin{tabular}{cc}
        \includegraphics[width=8.3cm]{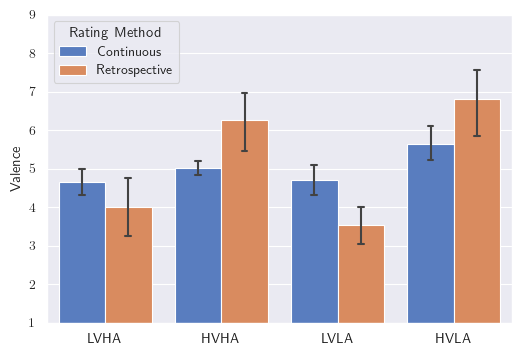} & 
        \includegraphics[width=8.3cm]{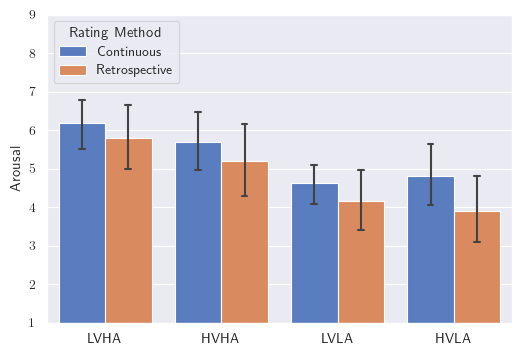}\\
        (a) Valence & (b) Arousal
    \end{tabular}    
    \caption{Mean ratings grouped by quadrant and rating method (whiskers representing the 95\% confidence interval).}
    \label{fig:fig_compare_values}
\end{figure*}

\subsection{Valence, Continuous versus Retrospective}
In group LVHA, the valence was lower in the retrospective rating than in the continuous rating. A Wilcoxon signed-rank test showed no significant difference ($z=1.71, p=0.086$), with a small positive effect size ($r=0.27$). In group HVHA, the valence was higher in the retrospective rating than in the continuous rating. A Wilcoxon signed-rank test revealed a significant difference ($z=-2.61, p=0.009$), with a moderate negative effect size ($r=-0.41$). In group LVLA, the valence was lower in the retrospective rating than in the continuous rating. A Paired $t$-test revealed a significant difference ($t=3.84, p=0.001$), with a strong positive effect size ($d=1.11$). In group HVLA, the valence was higher in the retrospective rating than in the continuous rating. A Wilcoxon signed-rank test revealed a significant difference ($z=-2.17,p=0.030$), with a moderate negative effect size ($r=-0.34$). The statistical tests for valence are displayed in Table \ref{table:valence_tests}. All valence ratings from participants are visualized, grouped by quadrant and rating method in Fig. \ref{fig:fig_compare_values}a.

\subsection{Arousal, Continuous versus Retrospective}
In group LVHA, the arousal was lower in the retrospective rating than in the continuous rating. A Wilcoxon signed-rank test showed no significant difference ($z=1.20, p=0.232$), with a small positive effect size ($r=0.18$). In group HVHA, the arousal was lower in the retrospective rating than in the continuous rating. A Paired $t$-test revealed no significant difference ($t=0.87, p=0.395$), with a small positive effect size ($d=0.24$). In group LVLA, the arousal was lower in the retrospective rating than in the continuous rating. A Paired $t$-test revealed no significant difference ($t=1.38, p=0.184$), with a small to moderate positive effect size ($d=0.30$). In group HVLA, the arousal was lower in the retrospective rating than in the continuous rating. A Paired $t$-test revealed no significant difference ($t=1.47, p=0.158$), with a moderate positive effect size ($d=0.45$). The statistical tests for arousal are displayed in Table \ref{table:arousal_tests}. All arousal ratings from participants are visualized, grouped by quadrant and rating method in Fig. \ref{fig:fig_compare_values}b.

\subsection{Evaluation of Rating Methods}
When asked to evaluate the different rating methods on a scale from 1 to 5, with 1 representing the lowest rating, the retrospective rating received an average rating of 3.8 ($SD=0.523$), compared to the continuous rating method, which received an average rating of 3.35 ($SD=1.089$). A Wilcoxon signed-rank test showed no significant difference ($z=27.0, p=0.097$) between these evaluations.

\section{Discussion}

\subsection{Correlation to Baseline Values}
Pearson's correlation coefficients show a strong correlation between valence ratings and a medium to strong correlation with arousal ratings, which indicates a valid emotional response to the video stimuli. A large difference between the raised values and the baseline values may be due to the video clips, which were shortened for this experiment to a consistent duration of one minute. This can lead to varying emotional responses depending on the scene chosen from the original video. Furthermore, it should be noted that the sample size was decreased by only using values from the retrospective rating method and by calculating means for each video instead of grouping by quadrant. Apart from these factors, the divergence could also be attributed to the remote setting, with the challenges of remote moderation, as one cannot adjust, show, or help participants in person. Due to these limitations, the correlation results must be viewed as an initial evaluation to have a glimpse into comparability with a lab experiment.

\subsection{Valence, Continuous versus Retrospective}
Apart from the LVHA quadrant, all the means between valence in continuous and retrospective were significantly different, with moderate to strong effect sizes. With three out of the four significant differences, there is evidence to reject the null hypothesis for \(H_1\), indicating that the two rating methods produce divergent results for emotion ratings. As a general trend, the values for the continuous method are more strongly centered around the mean value, and the retrospective values are more scattered.

One reason could be the interface that was used: a 9-point Likert scale (Fig. \ref{fig:fig_retrospective}), versus a 2D plane (Fig. \ref{fig:fig_continuous}). The difference in the rating interface seems to be valid and could cause these deviations but seems more unlikely when looking at the data collected for arousal ratings. Because the same method was used for both emotion dimensions, a similar tendency should be observable. In the absence of this effect, the spread cannot be attributed solely to the graphical rating interface.

The other aspect is the moment of rating, retrospective rating, and post-condition, compared to continuous ratings during the video that were then averaged. Interestingly, the effect sizes are negative for high valence videos (HVLA, HVHA), indicating that the mean valence was rated higher in retrospect. The opposite effect is observable for low valence videos (LVLA, LVHA), where the effect size is positive, indicating that the mean valence rating was lower in retrospect. This result supports the effect of an overestimation of valence in the post-questionnaires, similar to what the peak-end rules predict. However, this dynamic occurs naturally when the values of the continuous ratings are more balanced towards the mean.

\subsection{Arousal, Continuous versus Retrospective}
For arousal ratings, no significant differences are found in any of the four quadrants, despite the differences in rating interface and rating time. Consequently, the null hypothesis for \(H_2\) cannot be rejected. The effect sizes are low to moderate positive, indicating a slightly higher rating for the continuous method. While this difference in means is not significant, a higher excitement while watching the video would be reasonable, especially when considering that participants were alerted to give a rating every 10 seconds. An overestimation effect cannot be observed for arousal, possibly because the arousal values are not centered around the mean other than for valence. This could indicate that post-experience biases have different dynamics for arousal.

\section{Conclusion}

This study explored the dynamics of remote testing by conducting an experiment on emotion elicitation. Different rating methods have been employed to gather valence and arousal ratings to investigate the variability in self-reported measurements. A moderate to strong correlation of emotion ratings between the baseline values and experimental data was observed. Although this is not proof of the feasibility of remote studies, it ensures that the 360° videos have a comparable emotional effect on participants. The differences can be attributed to the small sample size, shortened stimuli, and difficulties in remote moderation, such as the inability to ensure optimal adjustment of the HMD. Additionally, the remote setting itself may have an impact on the UX. Specifically, the emotions experienced by participants may vary depending on whether they are located in a lab environment or their private home, potentially affecting the arousal and valence ratings during the experiment. While conducting a remote study proved to be a valuable method to continue research during times of pandemic, it was accompanied by additional considerations. In particular, the shipment of VR devices to the participant's home may not be feasible for studies with high participant counts. Nonetheless, the experiment procedure itself and remote moderation could be executed without major problems. This suggests that remote studies with user-owned devices could be an efficient approach unless the study requires participants who are inexperienced with VR.

When comparing the emotion ratings from the continuous and retrospective methods, different results were observed for valence and arousal. There is a significant difference in valence between the rating methods for the groups HVLA, HVHA and LVLA, supporting \(H_1\). In the retrospective ratings, the valence is more pronounced, corresponding to the original valence of the stimuli. This is congruent with the assumptions of the peak-end rule that an emotion rating can be overestimated if surveyed in retrospect. However, this cannot be solely attributed to a psychological effect; the lesser spread of continuous ratings can be caused by other differences in the rating method (e.g., interface, data processing, and interruptions). Unlike valence, the arousal ratings showed no significant differences, but a slight decrease in the means in retrospect, regardless of quadrant. Consequently, insufficient evidence supports \(H_2\). The different test results for valence and arousal reduce the possibility of influencing factors stemming from the rating methods themselves and suggest that there may be different dynamics in place for arousal and valence.

Whether the retrospective or continuous rating method is closer to the \enquote{real} emotions of participants cannot be determined. The subjective evaluation of the rating methods showed no significant difference, but participants tended to prefer the retrospective method.  However, it was found that rating methods can produce different outcomes for valence and arousal. Further, the experiment illustrates the feasibility of conducting remote studies by providing HMDs for participants and instruction over video conferencing. In summary, our study provides valuable insights into the complexities of assessing emotions in remote VR experiences, highlighting the importance of choice in rating methods and acknowledging the unique challenges posed by remote settings.

\subsection{Limitations and Future Work}

Some limitations of this study should be considered when interpreting its results. First, the impact of participants being in their homes instead of in a laboratory has not yet been explored and cannot be investigated using this experimental design. However, replication of this study in a laboratory environment could provide details on the influencing factors of the remote setting, such as the effect of the home environment on emotional reactions. The sample size of 20 participants, mostly male students, had limited generalizability. Future studies should include larger and more diverse groups of participants to eliminate the necessity of grouping the video stimuli by quadrant and possible gender bias. With enough equally distributed ratings for each video clip, a more precise comparison would be possible. The 60-second video clips used in this study cannot be regarded as equal to the original video source, which leads to limited comparability. Future studies investigating the impact of different rating methods should introduce more conditions to isolate factors such as the interface and timing for more meaningful insights. For continuous rating methods, different approaches to data processing should be explored. Instead of averaging all values, more elaborate calculations could lead to better results. Therefore, by further exploring these future paths, VR research may progress and better comprehend the remote testing of VR experiences and subjective emotion elicitation.


\bibliographystyle{IEEEtran}
\bibliography{IEEEabrv, main}
\nocite{*}
\end{document}